\documentclass[runningheads]{llncs}
\usepackage[T1]{fontenc}
\usepackage[english]{babel}
\usepackage{graphicx}
\usepackage{hyperref}
\hypersetup{
    pdftitle={Verifying ε-Differential Privacy Bit by Bit},
    colorlinks,
    linkcolor={red!50!black},
    citecolor={blue!50!black},
    urlcolor={blue!80!black},
}

\usepackage{xcolor}

\usepackage{amsmath}
\usepackage{amssymb}
\usepackage{thmtools}
\usepackage{mathtools}
\usepackage{breqn}
\usepackage{xcolor}
\usepackage{xspace}
\usepackage{booktabs}
\usepackage[noend]{algpseudocode}
\usepackage{algorithm}
\usepackage[numbers]{natbib}
\usepackage{cleveref}

\newlength\AlgIndent
\newcommand*\Let[2]{\State #1 $\gets$ #2}

\newcommand*\dom[1]{\mathrm{dom}\left( #1 \right)}

\algrenewcommand\algorithmicindent{0.75em}%
\algnewcommand\algorithmicswitch{\textbf{switch}}
\algnewcommand\algorithmiccase{\textbf{case}}
\algnewcommand\algorithmicassert{\texttt{assert}}
\algnewcommand\Assert[1]{\State \algorithmicassert(#1)}%

\crefformat{section}{§#2#1#3}
\Crefformat{section}{§#2#1#3}
\crefrangeformat{section}{§#3#1#4--#5#2#6}
\Crefrangeformat{section}{§#3#1#4--#5#2#6}
\crefmultiformat{section}{§#2#1#3}{ and~§#2#1#3}{,~§#2#1#3}{ and~§#2#1#3}
\Crefmultiformat{section}{§#2#1#3}{ and~§#2#1#3}{,~§#2#1#3}{ and~§#2#1#3}

\newcommand{\SYSTEM}{\textsc{FPlinko}\xspace}
\newcommand{\plinko}{\textsc{Plinko}\xspace}
\newcommand{\klee}{\textsc{KLEE}\xspace}
\newcommand{\approxln}{\textsc{Approx}\xspace}
\newcommand{\rlibm}{\textsc{RLibm}\xspace}
\newcommand{\coremath}{\textsc{core-math}\xspace}
\newcommand{\R}{\mathcal{R}}
\newcommand{\M}{\mathcal{M}}
\newcommand{\e}{\varepsilon}
\newcommand{\pc}{\varphi}
\newcommand{\pWhile}{\mathbf{pWhile}}
\newcommand{\Lap}{\mathrm{Lap}}
\newcommand{\emin}{e_{\mathrm{min}}}
\newcommand{\emax}{e_{\mathrm{max}}}
\newcommand{\F}{\mathbb{F}}
\newcommand{\Farb}{\F_p^w}
\newcommand{\half}{\F_{11}^{5}}

\newcommand{\double}{\F_{53}^{11}}
\newcommand{\LN}{\mathtt{LN}}
\newcommand{\sm}{\M_{\mathrm{SM}}}

\newcommand*\BitOr{\mathbin{|}}
\newcommand*\ShiftLeft{\ll}

\renewcommand{\exp}[1]{\mathrm{exp}\left(#1\right)}
\def\CC{{C\nolinebreak[4]\hspace{-.05em}\raisebox{.4ex}{\tiny\bf ++}}}
\DeclareMathOperator*{\lcm}{lcm}
\DeclarePairedDelimiter\abs{\lvert}{\rvert}

\definecolor{blue}{RGB}{0, 114, 178}
\definecolor{orange}{RGB}{213, 94, 0}

\NewDocumentCommand{\NewRestatableEnvironment}{m m}{
    \NewDocumentEnvironment{#1}{m +b}{%
        \begin{restatable}{donothing}{##1}
        \begin{#2}
        ##2
        \end{#2}
        \end{restatable}%
        }{\unskip\ignorespacesafterend}
    }

\NewRestatableEnvironment{restatableequation}{equation}

\begin{document}
\title{Verifying $\e$-Differential Privacy Bit by Bit}
%
%
\author{Zachary Susag\inst{1}\orcidID{0000-0002-3981-9529} \and
Justin Hsu\inst{1}\orcidID{0000-0002-8953-7060}}

\authorrunning{Susag and Hsu}
%
\institute{Cornell University, Ithaca NY 14850, USA\\
\email{\{zjs,justin\}@cs.cornell.edu}}
\maketitle              
\begin{abstract}
We propose a symbolic execution method for verifying $\e$-differential privacy of randomized programs that use IEEE~754 floating-point arithmetic.
In contrast to existing approaches, which reason over the real numbers, our method computes the exact floating-point support of the output distribution and the rational probability mass of each output.
Our technique augments the probabilistic symbolic execution framework of~\citet{susag2022} with bit-precise reasoning about floating-point arithmetic, and encodes the resulting differential privacy obligation as a logical query that can be discharged either by an off-the-shelf SMT solver or by an embarrassingly parallel exhaustive enumeration procedure.
We implement our method in a prototype tool,~\SYSTEM, on top of the~\klee symbolic execution engine and the Z3 SMT solver.
We evaluate~\SYSTEM on half-precision implementations of the Laplace and snapping mechanisms written in~\CC.
Our tool not only reproduces Mironov's reconstruction attack against the Laplace mechanism but also verifies that the snapping mechanism satisfies $\e$-differential privacy.
To the best of our knowledge, these are the first formal analyses of $\e$-differential privacy for floating-point implementations of the Laplace and snapping mechanisms.

\keywords{Differential privacy \and Floating-point arithmetic \and Program verification \and Symbolic execution \and Satisfiability modulo theories (SMT).}
\end{abstract}

\section{Introduction}
\label{sec:intro}

Every day, vast amounts of data are collected and aggregated into massive databases, including medical records, location metadata, and financial information.
Much of this data could drive scientific and medical research, but much of it cannot be safely released due to privacy concerns.
\textit{Differential privacy}~\citep{dwork2006} has established itself as the de facto standard for formalizing privacy over statistical queries; it has been adopted by the U.S.~Census Bureau to protect demographic data in the 2020 Census~\citep{abowd2018} and by Apple to learn aggregate user behaviors (e.g., browsing patterns, emoji usage)~\citep{cormode2018}.
In a nutshell, a randomized program is differentially private if, on any two input databases differing in a single person's private data, the program's output distributions are \textit{almost} the same; intuitively, a private program should not depend on, or reveal too much about, any single individual's record.
Privacy is modeled \textit{quantitatively}, by bounding how much the output distribution can change between two adjacent databases by a small constant $e^{\e}$, where $\e$ is a parameter controlling the tradeoff between privacy and accuracy.

Differential privacy is typically achieved by adding some random statistical noise to a numerical query's result: too much noise destroys accuracy and utility, while too little can violate privacy.
Crafting such algorithms correctly is non-trivial, and bugs have been discovered in commonly used mechanisms~\citep{lyu2017}.
A range of systems provide private building blocks and methods for implementing differential privacy in practice, including PINQ~\citep{mcsherry2009}, Airavat~\citep{roy2010}, and Fuzz~\citep{reed2010}, alongside open-source libraries of common differential privacy algorithms such as the OpenDP library~\citep{shoemate2026}, IBM's Diffprivlib~\citep{diffprivlib}, and Google's differential privacy library~\citep{2026a}.
While these systems differ in their implementation languages, target operating systems, and programming models, they share essentially the same noise-generation mechanism central to differential privacy.

Unfortunately, as~\citet{mironov2012} showed, many implementations of differentially private algorithms share a common simplification that leaves them vulnerable to attack: they reason about differential privacy over the real numbers, rather than over the floating-point numbers that actual implementations manipulate.
Mironov showed that the textbook Laplace mechanism, when implemented in double-precision floating-point arithmetic, admits a concrete reconstruction attack that breaks $\e$-differential privacy by exploiting the porous output distribution induced by finite precision and rounding.
At the time of publication, his attack succeeded against every publicly available general-purpose differential privacy library that was evaluated, including several mentioned above~\citep{mcsherry2009,roy2010,reed2010,mironov2012}.
Workarounds exist, such as Ilvento's base-2 exponential mechanism~\citep{ilvento2020} and Mironov's own snapping mechanism~\citep{mironov2012}, but the fact that so many independently developed implementations are susceptible to the same attack underscores the need to verify the differential privacy guarantees of concrete implementations of differentially private mechanisms.

\paragraph{Challenges.}
Verifying floating-point implementations of $\e$-differentially private mechanisms poses three distinct challenges that, to our knowledge, no existing approach addresses simultaneously:
\begin{itemize}
\item \textbf{Floating-point semantics.} As Mironov demonstrated, differential privacy guarantees proven over the real numbers do not automatically transfer to floating-point implementations.
  A sound analysis must track the exact floating-point support of the output distribution, including holes induced by sampling and rounding, and accommodate the semantically complex behaviors typical of these implementations: bitwise manipulation of binary interchange formats, classification of special values (e.g., subnormals), and calls to transcendental functions.
  Most existing verification approaches abstract these details away (\Cref{sec:rw}).
\item \textbf{Inferring exact probabilities for all inputs.} Verifying differential privacy requires reasoning about the exact output distribution of a randomized program.
  Statistical techniques can effectively detect violations~\citep{bichsel2021,zhang2020}, but cannot prove that a mechanism is differentially private for all inputs.
\item \textbf{Mainstream languages.} We are primarily interested in verifying differentially private mechanisms found in the wild and written in mainstream languages (e.g., C, Python).
  Several promising verification techniques rely on specialized static type systems~\citep{reed2010,gaboardi2013,barthe2015}; however, these systems do not exist in mainstream programming languages, creating a barrier to adoption.
\end{itemize}

\paragraph{Our approach.}
We propose an automatic probabilistic symbolic execution-based technique for verifying $\e$-differential privacy of real-world randomized programs implemented in floating-point arithmetic.
Prior symbolic execution approaches to differential privacy~\citep{wang2020,zhang2020,farina2021} either do not support floating-point operations or can only disprove differential privacy by synthesizing a counterexample.
Our work builds on~\citet{susag2022}, who developed a symbolic execution method for randomized programs that verifies quantitative correctness properties bounding the probability of some event.
Their method models sampling statements by introducing a new kind of \textit{probabilistic} symbolic variable that represents the result of sampling from a known distribution; the resulting symbolic formulas encode the probability of each program path, parameterized by unknown program inputs.

We adapt~\citet{susag2022}'s approach to verify $\e$-differential privacy by (i) adding support for reasoning about floating-point arithmetic, and (ii) encoding $\e$-differential privacy as a logical query over per-path probabilities.
This strategy lets us automatically prove that a given randomized program using IEEE~754~\citep{ieee754-2008} floating-point arithmetic is $\e$-differentially private for all inputs; otherwise, a witnessing counterexample is generated.

\paragraph{Contributions and outline.}

After reviewing background material (\Cref{sec:background}), we present our main contributions:
\begin{itemize}
\item An automatic symbolic-execution-based technique for verifying $\e$-differential privacy guarantees of randomized programs that use IEEE~754 floating-point arithmetic (\Cref{sec:meat}).
\item An optimized implementation of our approach called~\SYSTEM~(\Cref{sec:impl}).
  \SYSTEM is built on the~\klee symbolic execution engine~\citep{cadar2008}, to which we add floating-point support in the style of~\citet{liew2017}, and the Z3 SMT solver~\citep{demoura2008}.
  Through~\klee,~\SYSTEM can target randomized programs written in any language that compiles to LLVM bitcode, while faithfully modeling program states on real hardware (e.g., overflowing arithmetic, arrays, floating-point rounding artifacts).
\item An experimental evaluation of our tool (\Cref{sec:eval}).
  We use~\SYSTEM to (i) generate concrete counterexamples to $\e$-differential privacy for the textbook Laplace mechanism implemented in floating-point arithmetic, and (ii) \textbf{provide, to the best of our knowledge, the first automated proof of $\e$-differential privacy for an implementation of the snapping mechanism}~\citep{mironov2012}.
  We also discuss factors affecting the performance of~\SYSTEM, such as the choice of natural logarithm implementation.
\end{itemize}

We conclude by discussing related work (\Cref{sec:rw}) and future directions (\Cref{sec:conclusion}).
\section{Background}
\label{sec:background}

\subsection{Differential Privacy}
\label{sec:dp}
Differential privacy~\citep{dwork2006} is a quantitative, statistical notion of data privacy for computations over databases.
Informally, a randomized program is differentially private if, on any two input databases differing only in a single person's private data, the program's output distributions are ``\textit{nearly} equal''; equivalently, the influence any one person's record can have on the output is bounded.

Formally, let a \textit{database} $x \in \mathcal{D}$ be a finite sequence of records drawn from some universe $D$.
Two databases $x,y \in \mathcal{D}$ are \textit{adjacent} if they differ in at most one record, written $(x,y) \in \Delta$, where $\Delta$ is a fixed adjacency relation.
Quantitatively, ``\textit{nearly} equal'' is defined in terms of $\e \geq 0$ as follows.
\begin{definition}[$\e$-Differential Privacy~\citep{dwork2006}]
  \label{def:dworkdp}
  A randomized mechanism $\M: \mathcal{D} \rightarrow \R$ is $\e$-\textit{differentially private} with respect to an adjacency relation $\Delta$ if for every $x,y \in \mathcal{D}$ where $(x,y) \in \Delta$ and for any subset $S \subseteq \R$,
  \begin{equation*}
    \Pr\left[\M(x) \in S\right] \leq e^{\e}\Pr\left[\M(y) \in S\right]
  \end{equation*}
\end{definition}

For the purposes of this work, we adopt an equivalent pointwise reformulation of~\Cref{def:dworkdp}, which requires that the inequality hold for every output $z \in \R$:
\begin{definition}[Pointwise $\e$-Differential Privacy~\citep{dwork2006}]
  \label{def:dp}
  A randomized mechanism $\M: \mathcal{D} \rightarrow \R$ is $\e$-\textit{differentially private} with respect to an adjacency relation $\Delta$ if for every $x,y \in \mathcal{D}$ where $(x,y) \in \Delta$ and for every $z \in \R$,
  \begin{restatableequation}{dpineq}
    \label{eq:dp}
    \Pr\left[\M(x) = z\right] \leq e^{\e}\Pr\left[\M(y) = z\right]
  \end{restatableequation}
\end{definition}
This alternative definition is convenient because we will later verify differential privacy by reasoning about one output at a time.

\subsubsection{Laplace Mechanism}
\label{sec:background_laplace}

Many techniques convert non-private numerical queries into $\e$-differentially private mechanisms.
Most, but not all~\citep{blocki2012,bhaskar2011}, perturb the output by adding noise sampled from a specific distribution to the exact result of the non-private query, calibrated to its \textit{sensitivity}.
The Laplace mechanism is the canonical primitive for releasing a real-valued query under $\e$-differential privacy: it adds noise from the Laplacian distribution $\Lap(\lambda)$ scaled by the sensitivity.
\begin{definition}[$\ell_1$-sensitivity]
  A deterministic real-valued function $f: \mathcal{D} \rightarrow \R$ has $\ell_1$-\textit{sensitivity} at most $\Delta_f$ if for all adjacent databases $x,y \in \mathcal{D}$:
  \begin{equation*}
    \Delta_f = \max_{\substack{x,y \in \mathcal{D}\\\abs{x - y} = 1}}\abs{f(x) - f(y)}.
  \end{equation*}
\end{definition}
\begin{definition}[Laplacian distribution]
  The (zero-mean) Laplacian distribution $\Lap(\lambda)$ with $\lambda > 0$ is the continuous probability distribution defined by the density function
  \begin{equation*}
    f_{\lambda}(x) \triangleq \frac{1}{2\lambda}\exp{-\frac{\abs{x}}{\lambda}}
  \end{equation*}
\end{definition}
Given a real-valued query $f: \mathcal{D} \rightarrow \R$ of sensitivity $\Delta_f$, the Laplace mechanism is defined by:
\begin{equation}
  \label{eq:laplace_mech}
  \M_{\Lap}(x, f(\cdot), \e) = f(x) + Y \qquad Y \sim \Lap(\Delta_f / \e).
\end{equation}

The Laplace mechanism is well-known to preserve $\e$-differential privacy over the real numbers~\citep{dwork2014,dwork2006}; an actual implementation, however, must compute $f(x) + Y$ in a finite-precision representation and produce $Y$ from a finite source of randomness.
The standard method of generating $Y$ is inverse transform sampling: given the inverse $F^{-1}(\cdot)$ of a probability distribution's cumulative distribution function (CDF) and a source of randomness $U$ sampled uniformly from $[0,1)$, compute $Y \sim F^{-1}(U)$.
Since the Laplace distribution can be viewed as the exponential distribution paired with a randomly chosen sign, $Y \sim \Lap(\lambda)$ can be generated by sampling a value from the exponential distribution with parameter $\lambda$ and an integer $Z$ uniformly from $\{0,1\}$~\citep{mironov2012}:
\begin{equation}
  \label{eq:lapdist1}
  Y \sim (2Z - 1)\cdot\lambda\ln(1-U).
\end{equation}
Alternatively, one can sample a single value $U$ from the uniform distribution over $[0,1)$~\citep{mironov2012}:
\begin{equation}
  \label{eq:lapdist2}
  Y \sim \mathrm{sign}(r)\cdot\lambda\ln(1 - 2\abs{r}), \text{ where } r \sim U - 0.5.
\end{equation}
As we explain in~\Cref{sec:mironov}, this seemingly innocuous sampling procedure is the source of the floating-point vulnerabilities that defeat the Laplace mechanism's $\e$-DP guarantee.

\subsection{Floating-Point Arithmetic}
\label{sec:fp}

\begin{table}[t]
  \caption{Parameters of the binary interchange formats of size up to 128 bits specified by the IEEE~754-2008 standard~\citep{ieee754-2008}.}
  \centering
  \begin{tabular}{lrrrr}\toprule
    \textbf{Name}&\textbf{binary16}&\textbf{binary32}&\textbf{binary64}&\textbf{binary128}\\\midrule
    $p$ & 11 & 24 & 53 & 113 \\
    $\emin$ & -14 & -126 & -1022 & -16382 \\
    $\emax$ & +15 & +127 & + 1023 & +16383 \\\bottomrule
  \end{tabular}
  \label{tab:ieee754}
\end{table}

We give a brief overview of floating-point arithmetic and introduce notation; we refer the reader to~\citet{goldberg1991} for a concise introduction and to~\citet{muller2018} as a current reference.

A floating-point number $x$ in a floating-point format $\mathbb{F} \subseteq \mathbb{R}$ has the form
\begin{equation}
  \label{eq:fp}
  x = (-1)^s\cdot m \cdot \beta^{e-p+1},
\end{equation}
where $\beta \in \{ b \in \mathbb{N} \mid b \geq 2\}$ is the \textit{radix}, $p \in \{ p' \in \mathbb{N} \mid p' \geq 2 \}$ is the \textit{precision}, $m \in \mathbb{N} \cap [0, \beta^p)$ is the \textit{significand}, $e \in \mathbb{Z} \cap [\emin, \emax]$ is the \textit{exponent}, and $s \in \{0,1\}$ is the \textit{sign} bit of $x$.
The IEEE~754 floating-point standard~\citep{ieee754-2008} defines several floating-point formats of varying precision; the main parameters of the binary interchange formats up to 128 bits are given in~\Cref{tab:ieee754}.
Each IEEE~754 format obeys $\emin = 1 - \emax$ and reserves $\log_2(\emax + 1)+1$ bits for the exponent and $p-1$ bits for the significand, in addition to the sign bit.
We use $\Farb$ to denote the IEEE~754 binary floating-point format with precision $p$ and $w$ bits reserved for the exponent; for example, the set of all half-precision floating-point numbers (binary16) is~$\half$.
The standard also defines encodings for $\pm0$ ($e = 0; m = 0$), $\pm\infty$ ($e = \emax; m = 0$), and not-a-number (NaN) ($e = \emax; m \neq 0$) values to represent invalid operations (e.g., $0/0$).

Because $\mathbb{F}$ is a finite-precision format, many real numbers cannot be represented exactly: for instance, $0.1$ has no exact binary64 representation.
Moreover, the result of most elementary arithmetic operations (e.g., addition, subtraction, multiplication, division) on floating-point numbers must be \textit{rounded} back to a representable value.
The IEEE~754 standard~\citep{ieee754-2008} mandates that the results of all basic arithmetic operations be \textit{exactly rounded}: the result is first computed at infinite precision and then rounded to the nearest floating-point number under one of five \textit{rounding modes}.
Throughout this paper we assume the IEEE~754 default round-to-nearest, ties-to-even mode unless stated otherwise.
We circle each operator to distinguish floating-point operations from their real-arithmetic counterparts: $\oplus, \ominus, \otimes, \oslash$ are floating-point analogues of addition, subtraction, multiplication, and division, respectively.

\subsection{Mironov's Attack}
\label{sec:mironov}

The textbook analysis of $\M_{\Lap}$ implicitly assumes that the noise $Y$ is drawn from the \emph{true} Laplace distribution over $\mathbb{R}$.
Practical implementations, however, must sample $U$ uniformly from a finite subset of floating-point numbers in $(0,1)$ and pass it through floating-point implementations of the natural logarithm and the surrounding arithmetic.
\citet{mironov2012} demonstrated that this semantic gap is not benign: it permits a concrete reconstruction attack that breaks $\e$-DP for any finite $\e$.
At the time of publication, all four publicly available general-purpose DP libraries that were evaluated were susceptible to the attack~\citep{mironov2012}.

The crux of the attack is that a uniformly drawn floating-point number in $(0,1)$ is \emph{not} uniform over $(0,1)$ in any meaningful sense.
Most implementations generate $U$ either by dividing a uniformly random integer from $[0, 2^n-1]$ by $2^n$ or by an equivalent significand-fill construction~\citep{goualard2020,goualard2022}; the resulting distribution is supported on at most $2^{n}$ values evenly spaced by $2^{-n}$, even though $\Farb$ contains many more representable values in $(0,1)$ near zero.
To compound matters, different languages' standard libraries and numerical packages adopt different sampling constructions.
For example, Java's \texttt{Random.nextDouble()} returns a uniformly random double-precision ($\double$) value in $[0,1)$ by drawing a random 53-bit unsigned integer and multiplying it by $2^{-53}$, whereas ISO C provides no standard function for random floats in $[0,1)$ and instead encourages dividing \texttt{rand()} by \texttt{RAND\_MAX+1}, where \texttt{RAND\_MAX} is only $2^{31} - 1$ in the GNU Compiler Collection (GCC) library~\citep{goualard2020}.
After $U$ is passed through the inverse-transform procedure and added to the true result of the target query $f(\cdot)$, the image of $\M_{\Lap}$ is an irregular subset of $\Farb$ whose spacing inherits both the $2^{-n}$ uniform spacing of $U$ and the geometric spacing of $\Farb$.
Even if $U$ is drawn uniformly from $\Farb \cap (0,1)$, transforming $U$ via~\cref{eq:lapdist1} or~\cref{eq:lapdist2} introduces artifacts of its own.

The result is a distinctively porous output distribution: there exist floating-point numbers $z \in \Farb$ for which
\[
  \Pr[\M_{\Lap}(x, f(\cdot), \e) = z] > 0
  \quad\text{but}\quad
  \Pr[\M_{\Lap}(y, f(\cdot), \e) = z] = 0
\]
on adjacent databases $(x,y) \in \Delta$, immediately violating~\Cref{eq:dp} for every $\e$.
Mironov uses these ``smoking-gun'' outputs as the basis for his attack: by issuing a small number of queries to $\M_{\Lap}$ and observing whether such an output materializes, an adversary can distinguish adjacent databases with non-trivial advantage.
In the same work,~\citet{mironov2012} also proposed the \textit{snapping mechanism} as an attack-resistant alternative; we revisit it as a case study in~\Cref{sec:sm}.

\section{Probabilistic Symbolic Execution of $\e$-DP Mechanisms}
\label{sec:meat}

Our goal is to automatically verify whether a mechanism $\M$, implemented as a randomized program with floating-point arithmetic and discrete sampling, preserves $\e$-differential privacy.
By the pointwise definition of $\e$-DP (\Cref{def:dp}), this requires checking, \textit{for all} adjacent inputs $(x,y) \in \Delta$ and \textit{every} output $z \in \R$:
\dpineq*

\noindent{}As discussed in~\Cref{sec:mironov}, a successful approach must compute the \textit{exact} floating-point support of the program's output distribution and the \textit{exact} probability of each output.
We adopt a probabilistic symbolic execution technique that lets us compute $\Pr\left[\M(x) = z\right]$ exactly for symbolic parameters $x$ and $z$ while preserving the bit-level semantics of floating-point arithmetic.

\Cref{sec:psefp} describes our language and reviews standard symbolic execution before introducing the probabilistic variant of~\citet{susag2022}, which we extend with bit-precise reasoning over IEEE~754 floating-point expressions and conditionals.
\Cref{sec:encoding} converts the resulting paths --- whose per-path probability expressions retain the shape introduced by~\citet{susag2022} --- into a bitvector encoding of $\e$-DP as a logical query, and~\Cref{sec:solving} describes two strategies for discharging it: a direct call to an SMT solver, and an exhaustive enumeration algorithm that externalizes the probability computations.

\subsection{Probabilistic Symbolic Execution for Floating-Point Programs}
\label{sec:psefp}

\subsubsection{Language}
\label{sec:lang}
We model mechanism implementations written in $\pWhile$, a core imperative probabilistic programming language, extended with floating-point arithmetic.
The statements of $\pWhile$ are described by the following grammar:
\[
  S := \mathtt{x} \gets e \mid \mathtt{x} \sim d \mid S_1 ; S_2 \mid \mathbf{if}~c~\mathbf{then~goto}~T \mid \mathbf{halt}
\]

The assignment statement $\mathtt{x} \leftarrow e$ binds the program variable $\mathtt{x}$ to the result of evaluating expression $e$, and the sampling statement $\mathtt{x} \sim d$ binds the program variable $\mathtt{x}$ to a random sample from a distribution $d$.
Control flow is handled by $S_1 ; S_2$, which sequences two statements $S_1$ and $S_2$, and $\mathrel{\mathbf{if}} c \mathrel{\mathbf{then~goto}} T$, which jumps execution to the statement referenced by the address/label $T$ if the guard $c$ holds, otherwise falling through to the next instruction.
We assume that high-level constructs like loops, function calls, and regular conditionals are compiled down to conditional jumps.
Finally, $\mathbf{halt}$ marks the end of a program's execution.

Expressions $e$ have one of three types: Booleans, finite-precision machine integers, or binary floating-point numbers $\F_p^w$.
Integer expressions are built from the usual fixed-width arithmetic operators (e.g., addition, subtraction, multiplication, division, modulus), bitwise operators, and relational operators; floating-point expressions are built from the corresponding floating-point operators $\oplus, \ominus, \otimes, \oslash$, together with standard comparisons, classification predicates for subnormal numbers, infinity, and NaN, casts between floating-point formats, and casts between machine integers and floating-point values.
Branch guards $c$ are Boolean-valued expressions and may freely mix integer and floating-point comparisons.
Arrays are also supported by our tool,~\SYSTEM, but for presentation we assume each array element is modeled by a distinct program variable.

Distribution expressions $d$ denote the \textit{discrete} distribution random samples should be drawn from.
Each distribution corresponds to a mathematically standard distribution (e.g., uniform or Bernoulli).
We interpret distributions as functions from \textit{integers} to $\mathbb{Q} \cap [0, 1]$, returning the \textit{rational} probability of each integer.
For example, $\mathsf{UniformInt}(\mathtt{1},\mathtt{6})$ selects a random integer between $1$ and $6$ inclusive.
Restricting the domain to integers is an implementation convenience, not a fundamental limitation: any finite, discrete distribution over a floating-point format with rational, computable per-element probabilities could be supported.
Our case studies use only uniform integer, Bernoulli, and finite uniform joint distributions, which together suffice to express the sampling procedures of both the Laplace (\Cref{sec:lm}) and snapping mechanisms (\Cref{sec:sm}).

\subsubsection{Symbolic Execution}
\label{sec:symex}
Symbolic execution~\citep{king1976} is a classic program analysis technique that explores the set of paths through a program by treating each input as a \emph{symbolic variable} and propagating symbolic expressions in place of concrete values.
Program state is modeled using a pair of data structures: a path condition $\varphi$, which is a conjunctive formula consisting of all the symbolic branch constraints that hold for a particular path, and an expression map $\sigma$, which maps program variables to symbolic expressions built from constants and symbolic variables.
Throughout, we refer to a path by its path condition $\pc$.

Symbolic execution begins by initializing $\sigma$ to map each program input to a fresh symbolic variable.
The algorithm then iterates through all reachable states, analyzing each instruction and updating $\pc$ and $\sigma$ as needed.
For an assignment $\mathtt{x} \gets e$, the program expression $e$ is first converted into a symbolic expression by applying $\sigma$.
We write $\sigma[e] = e_{sym}$ for the symbolic expression obtained by replacing each program variable in $e$ with its corresponding symbolic expression in $\sigma$.
Upon encountering a branch $\mathrel{\mathbf{if}} c \mathrel{\mathbf{then~goto}} T$, execution \textit{forks}: the true branch is explored with path condition $\pc \wedge c_{sym}$ and the false branch with $\pc \wedge \neg c_{sym}$.

\subsubsection{Probabilistic Symbolic Execution}
\label{sec:pse}

We now explain how sampling instructions are handled.
\citet{susag2022} distinguish two types of symbolic variables: \textit{universal} variables $\alpha_1, \alpha_2, \ldots$, which represent unknown program inputs and are universally quantified in the final verification query, and \textit{probabilistic} variables $\delta_1, \delta_2, \ldots$, which stand in for the result of a random draw from a known distribution.
Upon encountering a sampling statement $\mathtt{x} \sim d$, a fresh probabilistic symbolic variable $\delta$ is introduced and $\sigma$ is updated so that $\sigma[\mathtt{x}] = \delta$.
We stress that $\delta$ is treated as a free variable over $d$'s support, not as an opaque constant: this is what lets us later compute branch and path probabilities symbolically rather than approximate them, e.g., via Monte Carlo sampling.

Two additional data structures track the probabilistic information accrued during symbolic execution:
\begin{itemize}
\item \textit{Distribution map} ($P$): Similar in spirit to $\sigma$, $P$ maps each probabilistic symbolic variable to the distribution expression it was originally sampled from.
\item \textit{Path probability} ($p$): For each path, we maintain a symbolic expression for the path probability $p$ over the universal symbolic variables. As we explain later, this expression contains no \textit{probabilistic} symbolic variables.
\end{itemize}

For mechanisms implemented in $\pWhile$, the probabilistic symbolic execution algorithm of~\citet[Algorithm 1]{susag2022} can be reused largely without modification.
Their algorithm never inspects the structure of the expression sublanguage directly, so extending it with floating-point arithmetic requires only that the decision procedure used to check path feasibility can reason about IEEE~754 floating-point semantics.
No new statement-level cases are needed, and the underlying data structures carry over.

\subsubsection{Computing Branch Probabilities}
\label{sec:branchprob}

\begin{algorithm}[t]
	\caption{\textsc{PBranch}: per-branch probability expression for a guard
$c_{\mathrm{sym}}$ under path condition $\pc$ and distribution map $P$.
Reproduced from~\citet[Algorithm 2]{susag2022}; the simplified form below
assumes uniform integer sampling distributions, and the general weighted case
is handled analogously.}
	\label{alg:pbranch}
	\begin{algorithmic}[1]
		\Function{PSEBranch}{$c_{sym}, \pc, P$}
		\Let{$(\delta_1,\ldots,\delta_n)$}{$\dom{P}$}
		\Let{$(d_1,\ldots,d_n)$}{$(P[\delta_1],\ldots,P[\delta_n])$} 
    \Let{$\mathcal{D}$}{$\dom{d_1} \times \cdots \times \dom{d_n}$}\label{line:D}
		\Let{$p_c$}{$\frac{\displaystyle{\sum_{(v_1,\ldots,v_n) \in \mathcal{D}}} [(c_{sym} \wedge \pc)\{\delta_1 \mapsto v_1,\ldots,\delta_n \mapsto v_n\}]}{\displaystyle{\sum_{(v_1,\ldots,v_n) \in \mathcal{D}}}[\pc\{\delta_1 \mapsto v_1,\ldots,\delta_n \mapsto v_n\}]}$}\label{line:branch_prob_compute}
    \Let{$p_c'$}{$1-p_c$}
		\State\Return{$(p_c, p_c')$}
		\EndFunction
	\end{algorithmic}
\end{algorithm}

As in traditional symbolic execution, probabilistic symbolic execution explores both branches at each branch statement.
Unlike traditional symbolic execution, however, the presence of discrete sampling statements means each branch has an associated probability, possibly referencing universal symbolic variables.
\textsc{PBranch} (\Cref{alg:pbranch}) constructs the path probability expressions for the true and false branches.
It takes as input the symbolic guard $c_{sym}$, the current path condition $\pc$, and the current distribution map $P$, and returns symbolic expressions $(p_c, p_c')$ encoding the probabilities of the true and false branches, respectively.
We present \textsc{PBranch} in the simplified setting where all samples are drawn from discrete uniform distributions; the general weighted and joint case is handled analogously and is supported by our implementation.

We give a brief, intuitive justification for the correctness of~\Cref{alg:pbranch}; the formal proof appears in~\citet{susag2022}.
When a branch statement is encountered, $\pc$ contains all constraints on the symbolic variables required to reach this point.
We can therefore view the probability of taking either branch as a \textit{conditional probability}: the probability that $c_{sym}$ holds given $\pc$, namely $\Pr[c_{sym} \mid \pc] = \Pr[c_{sym} \wedge \pc]/\Pr[\pc]$.

Each probabilistic symbolic variable $\delta$ is mapped to exactly one distribution $d$.
If $P$ contains probabilistic symbolic variables $\delta_1, \ldots, \delta_n$ with distributions $d_1, \ldots, d_n$, the set of all assignments to $\delta_1, \ldots, \delta_n$ is $\mathcal{D}$.
Under our simplifying assumption that all distributions are uniform, each assignment $(v_1,\ldots,v_n) \in \mathcal{D}$ has equal weight.
Thus, instead of computing $\Pr[c_{sym} \mid \pc]$ as a ratio of probabilities, we encode it as the ratio of the number of assignments from $\mathcal{D}$ that satisfy $c_{sym} \wedge \pc$ and $\pc$, as shown on~\cref{line:branch_prob_compute} of~\Cref{alg:pbranch}, using Iverson brackets $[\cdot]$, where $[Q] = 1$ if $Q$ is true and 0 otherwise.
$p_c$ is not a constant but a \textit{symbolic expression} possibly containing free universal symbolic variables; equivalently, $p_c$ is a formula for $\Pr[c_{sym} \mid \pc]$ parameterized by those variables.
The last two lines of~\Cref{alg:pbranch} use the fact that conditional probabilities of complementary events sum to 1.

Given these branch probabilities, the probability of an entire path follows directly: since $p_c$ is the conditional probability of $c_{sym}$ given $\pc$ and $p$ is the probability of $\pc$ holding, the standard rule $\Pr[A \wedge B] = \Pr[A \mid B]\cdot\Pr[B]$ gives $\Pr[c_{sym} \wedge \pc] = p_c \cdot p$ and $\Pr[\neg c_{sym} \wedge \pc] = (1 - p_c) \cdot p = p_c' \cdot p$.

More generally, still under the uniform-distribution assumption, the probability expression for a complete path condition $\pc_i$ is:
\begin{equation}
  \label{eq:pathprob}
  p_{i} \triangleq \frac{\displaystyle\sum_{(v_1, \ldots, v_n) \in \mathcal{D}}[\pc_i\{\delta_1 \mapsto v_1, \ldots, \delta_n \mapsto v_n\}]}{\abs{\mathcal{D}}}
\end{equation}
General weighted distributions are handled by replacing the unit-weighted count in $\abs{\mathcal{D}}$ with the joint product of per-element weights.

\subsection{Encoding $\e$-Differential Privacy as a Logical Query}
\label{sec:encoding}

Recall that our original goal is to prove that a mechanism $\M$ preserves $\e$-differential privacy, which requires reasoning about the following inequality over all adjacent input pairs $(x,y) \in \Delta$ and all outputs $z$:
\dpineq*

\noindent{}We now transform the path and probabilistic information produced by the probabilistic symbolic execution of~\Cref{sec:psefp} into a logical query that encodes the pointwise definition of $\e$-DP (\Cref{eq:dp}) and is dischargeable by off-the-shelf SMT solvers.

Suppose probabilistic symbolic execution of $\M(x)$, given input $x$ as a universal symbolic variable, produces a finite collection of feasible paths $\pc_1, \ldots, \pc_n$, along with probability expressions $p_1, \ldots, p_n$ and expression maps $\sigma_1, \ldots, \sigma_n$.
Suppose further that each $\sigma_i$ contains a distinguished variable \texttt{out}, storing the output of $\M(x)$ along path $\pc_i$.
Pairing $\sigma_i[\mathtt{out}]$ with $p_i$ yields a formula for the probability that $\M(x)$ follows path $\pc_i$ \textit{and} produces a particular output $z$, represented by a new universal symbolic variable: $p_i \cdot [\sigma_i[\mathtt{out}] = z]$.
The total probability $\Pr[\M(x) = z]$ is then the sum over all $n$ paths:
\begin{equation}
  \label{eq:mechoutputprob}
  \Pr[\M(x) = z] = \sum_{i=1}^n p_i \cdot [\sigma_i[\mathtt{out}] = z].
\end{equation}
We stress that $\Pr[\M(x) = z]$ is a \textit{symbolic expression} encoding the probability that $\M$ produces the symbolic output $z$ given symbolic input $x$.
For presentation we show only two universal symbolic variables $x$ and $z$; the implementation of $\M$ may contain arbitrarily many.
The symmetric probability $\Pr[\M(y) = z]$ can be obtained by substituting $y$ for $x$ in $\Pr[\M(x) = z]$, i.e., $\Pr[\M(y) = z] = \left(\Pr[\M(x) = z]\right)\{x \mapsto y\}$.
We can then build a query of the following form: for all $(x,y) \in \Delta$ and for all outputs $z \in \Farb$,
\begin{equation}
  \label{eq:dpquery}
  \sum_{i=1}^n p_i \cdot [\sigma_i[\mathtt{out}] = z] \leq e^{\e} \cdot \left(\sum_{i=1}^n p_i \cdot [\sigma_i[\mathtt{out}] = z]\right)\{x \mapsto y\}.
\end{equation}

\subsubsection{Avoiding Mathematical Integers and Reals}
\label{sec:bv_encoding}

Discharging~\Cref{eq:dpquery} directly would require the decision procedure to support a theory rich enough to multiply the sum in~\Cref{eq:mechoutputprob} by the irrational constant $e^{\e}$ and to compare two such products as exact rationals or real numbers.
Combining such a theory with the theories of fixed-size bitvectors, floating-point arithmetic, and arrays that we already need for sound modeling of the mechanism would require a non-trivial theory combination, and in practice we have observed significant performance degradation when mixing the SMT-LIBv2~\citep{smtlib} \texttt{Real} theory with floating-point and bitvector reasoning.
We avoid this entirely by encoding $\Pr[\M(x) = z]$ and $\Pr[\M(y) = z]$ as ratios of fixed-width bitvectors and the DP inequality as a single bitvector inequality.

\paragraph{Common-denominator elimination.}
Because the $n$ paths $\pc_1, \ldots, \pc_n$ are produced by symbolically executing a \emph{single} mechanism $\M$, the per-path denominators in~\Cref{eq:pathprob} depend only on the structure of $\M$, not on the inputs.
Since both $\Pr[\M(x) = z]$ and $\Pr[\M(y) = z]$ are obtained by symbolically executing the same program on different symbolic inputs $x$ and $y$, the two probability expressions share a common denominator
\begin{displaymath}
  D \triangleq \displaystyle\lcm_{1 \leq i \leq n}\, \abs{\mathcal{D}_i},
\end{displaymath}
where $\mathcal{D}_i$ is the joint domain of the probabilistic symbolic variables free in $\pc_i$.
After multiplying by $D$, each probability is represented by an integer-valued weight
\begin{equation}
  \label{eq:weight}
  \begin{split}
    &D \cdot \Pr[\M(x) = z] \\&= \sum_{i=1}^n \frac{D}{\abs{\mathcal{D}_i}}
    \sum_{(v_1, \ldots, v_k) \in \mathcal{D}_i}
    \mspace{-9mu}
    \left[
    (\varphi_i \wedge \sigma_i[\mathtt{out}] = z)
    \{\delta_1 \mapsto v_1, \ldots, \delta_k \mapsto v_k\}
    \right],
  \end{split}
\end{equation}
which is a sum of Iverson brackets and therefore expressible directly as a fixed-size bitvector.
We thus multiply both sides of~\Cref{eq:dpquery} by $D$ before attempting to prove the query.

\paragraph{Rational approximation of $e^{\e}$.}
The remaining obstacle is the irrational privacy parameter $e^{\e}$.
Our solution is simple: approximate $e^{\e} \approx p/q$ with a sufficient amount of precision, where $p$ and $q$ are both fixed-size bitvectors.
We require $p/q \geq e^{\e}$ to ensure that any counterexample to~\Cref{eq:bv_dpquery} is also a counterexample to the original inequality with the exact $e^{\e}$.
We can then multiply both sides of~\Cref{eq:dpquery} again by the denominator $q$, resulting in a pure bitvector inequality:

\begin{equation}
  \label{eq:bv_dpquery}
  \begin{split}
    &q \cdot \left(\sum_{i=1}^n \frac{D}{\abs{\mathcal{D}_i}}
    \sum_{(v_1, \ldots, v_k) \in \mathcal{D}_i}
    \mspace{-20mu}
    \left[
    (\varphi_i \wedge \sigma_i[\mathtt{out}] = z)
    \{\delta_1 \mapsto v_1, \ldots, \delta_k \mapsto v_k\}
    \right]\right) \leq \\ &p \cdot \left(\sum_{i=1}^n \frac{D}{\abs{\mathcal{D}_i}}
    \sum_{(v_1, \ldots, v_k) \in \mathcal{D}_i}
    \mspace{-20mu}
    \left[
    (\varphi_i \wedge \sigma_i[\mathtt{out}] = z)
    \{x \mapsto y, \delta_1 \mapsto v_1, \ldots, \delta_k \mapsto v_k\}
    \right]\right),
  \end{split}
\end{equation}

\paragraph{Choosing a safe bitvector width.}
The terms in~\Cref{eq:bv_dpquery} must be evaluated in a bitvector whose width is large enough to avoid overflow for the products $q \cdot D \cdot \Pr[\M(x, f(\cdot), \e) = z]$ and $p \cdot D \cdot \Pr[\M(y, f(\cdot), \e) = z]$.
By construction, each Iverson bracket in~\Cref{eq:weight} contributes at most $1$ to the unscaled inner sum, so the inner sum is bounded above by $\abs{\mathcal{D}_i}$, and each per-path scaled contribution to the probability mass is bounded above by $D$.
Taking the sum over all $n$ paths yields the bound $D \cdot \Pr[\M(x, f(\cdot), \e) = z] \leq n \cdot D$, and applying the $\max(p, q)$ coefficient from~\Cref{eq:bv_dpquery} gives an upper bound of $\max(p, q) \cdot n \cdot D$ on every term that appears in the cross-multiplied inequality.
We therefore pick the bitvector width
\begin{equation}
  \label{eq:probwidth}
  w \triangleq \left\lceil\log_2\!\left(\max(p, q) \cdot n \cdot D + 1\right)\right\rceil,
\end{equation}
which is the smallest width in which $\max(p, q) \cdot n \cdot D$ can be represented as an unsigned integer.
Note that $w$ can be computed statically since all distributions are discrete and known at compile time.

\paragraph{Sensitivity and adjacency constraints.}
The final piece of the query consists of constraints on the inputs $x$ and $y$ enforcing $(x,y) \in \Delta$.
Both case studies in~\Cref{sec:eval} take a single floating-point value as input (which we assume is the result of applying a non-private query $f(\cdot)$ to some database), so we focus on this case here.
The adjacency relation $\Delta$ then reduces to a sensitivity bound of $1$ on the input, which we encode directly as
\begin{equation}
  \label{eq:adj_constraints}
  x \neq y \wedge \abs{x \ominus y} \leq_{\F} 1.0,
\end{equation}
where $\leq_{\F}$ denotes the floating-point less-than-or-equal operator.
Encoding the sensitivity bound in floating-point arithmetic, rather than as a real-valued constraint, ensures that any counterexample inputs $(x, y)$ returned by the solver are themselves bit-precise and can be replayed against the concrete implementation.

\paragraph{Putting it together.}
The full logical query consists of declarations of the universal symbolic variables $x, y, z$ in the appropriate floating-point sort, the adjacency constraints (\Cref{eq:adj_constraints}), and the negated DP inequality (\Cref{eq:bv_dpquery}):
\begin{equation}
  \label{eq:full_query}
  \begin{split}
    \exists &x, y, z \in \Farb ~:~ x \neq y \wedge \abs{x \ominus y} \leq_{\F} 1.0 \\
    &\wedge p_x = \sum_{i=1}^n \frac{D}{\abs{\mathcal{D}_i}}
    \sum_{(v_1, \ldots, v_k) \in \mathcal{D}_i}
    \mspace{-20mu}
    \left[
    (\varphi_i \wedge \sigma_i[\mathtt{out}] = z)
    \{\delta_1 \mapsto v_1, \ldots, \delta_k \mapsto v_k\}
      \right] \\
    &\wedge p_y = p_x\{x \mapsto y\} \\
    &\wedge q \cdot p_x > p \cdot p_y
  \end{split}
\end{equation}
where $p_x$, $p_y$, $D$, $\abs{D_i}$, $p$, $q$, and all Iverson-bracketed terms are all $w$-bit integers.
This query is unsatisfiable exactly when $\M$ is $\e$-DP under the rational upper bound $p/q$ for $e^{\e}$.
A satisfying assignment yields a counterexample: a triple $(x, y, z)$ together with the concrete probabilities $\Pr[\M(x) = z]$ and $\Pr[\M(y) = z]$ witnessing the violation.

The construction of the probability expressions inherits its soundness from prior work: for every feasible path, the symbolic expression produced by~\Cref{alg:pbranch} agrees with the true measure of $\Pr[\pc_i]$ under the joint distribution of the probabilistic symbolic variables, as established by~\citet[Theorems 3.2--3.4]{susag2022}.
Our floating-point extension to the expression sublanguage of $\pWhile$ does not modify any of the rules underpinning those theorems.

\subsection{Solving the Logical Query}
\label{sec:solving}

We support two interchangeable strategies for solving the query.
Both operate on the same logical query and either report a genuine counterexample or certify that the target mechanism is truly $\e$-differentially private.
We defer the quantitative comparison to~\Cref{sec:eval}.

\subsubsection{Direct Discharge}
\label{sec:solving_direct}
The first strategy is the most direct: serialize the query as an SMT-LIB script and hand it to an off-the-shelf SMT solver, such as Z3~\citep{demoura2008}.
The solver bears full responsibility for searching the input space and evaluating the symbolic probability expressions.
This strategy is attractive for its simplicity, requiring no auxiliary machinery beyond assembling the query itself.

\subsubsection{Exhaustive Enumeration}
\label{sec:solver_enum}

\begin{algorithm}[t]
  \caption{Exhaustive enumeration of adjacent inputs and mechanism outputs.
    The procedure iterates over every adjacent input pair $(x, y) \in \Delta$
    and every reachable output $z \in \Farb$, computes the probabilities
    $\Pr[\M(x) = z]$ and $\Pr[\M(y) = z]$ by
    directly evaluating the probability
    expressions of~\Cref{eq:mechoutputprob} on the concrete input/output
    pairs, and checks whether the bitvector form of the DP
    inequality~\Cref{eq:bv_dpquery} is violated. It returns either a concrete
    counterexample to $\e$-DP or reports that $\M$ is $\e$-differentially
    private.}
  \label{alg:exhaustive}
  \begin{algorithmic}[1]
    \Function{Enumerate}{$\Delta, \Farb, p, q$}
      \For{$(x, y) \in \Delta$}
        \For{$z \in \Farb$}
          \Let{$p_x$}{\textsc{Evaluate}$(\Pr[\M(x) = z])$}
          \Let{$p_y$}{\textsc{Evaluate}$(\Pr[\M(y) = z])$}
          \If{$q \cdot p_x > p \cdot p_y$}
            \State \Return $(x, y, z, p_x, p_y)$
              \Comment{counterexample to $\e$-DP}
          \EndIf
        \EndFor
      \EndFor
      \State \Return \textsc{Verified} \Comment{$\M$ is $\e$-DP}
    \EndFunction
  \end{algorithmic}
\end{algorithm}

The probability expressions in~\Cref{eq:full_query} can be very large sums of bitvector terms, especially for mechanisms with many paths or those that sample from distributions with large domains.
In our experience, evaluating these expressions often dominates the cost of direct discharge.
Our second strategy, shown in~\Cref{alg:exhaustive}, \textit{decouples} input-space search from probability evaluation.
Rather than relying on an external solver, we iterate through every adjacent input pair $(x,y) \in \Delta$ and every possible mechanism output $z$, computing the probabilities $\Pr[\M(x) = z]$ and $\Pr[\M(y) = z]$ by directly evaluating the expressions on the concrete pairs $(x,z)$ and $(y,z)$.
This computation is embarrassingly parallel, as we discuss further in~\Cref{sec:impl}.
After computing the probabilities, we check whether~\Cref{eq:bv_dpquery} is violated; if so, we have a concrete counterexample, and otherwise we continue.
If no counterexample is found before exhausting all pairs and outputs,~\Cref{alg:exhaustive} reports that $\M$ is $\e$-differentially private.

\section{Implementation}
\label{sec:impl}

We have implemented our approach in a tool called~\SYSTEM by extending the publicly available probabilistic verifier~\plinko~\citep{susag2022}.
\SYSTEM is written in Rust and consists of approximately 15,000 lines of code.
It is built on top of the~\klee symbolic execution engine~\citep{cadar2008} and uses the Z3 theorem prover~\citep{demoura2008}.
Currently,~\SYSTEM targets mechanisms implemented in C/\CC; however, because~\klee performs symbolic execution on LLVM~\citep{lattner2004} bitcode,~\SYSTEM can in principle analyze any program written in a language that compiles to LLVM intermediate representation (IR).
All that is needed to support other high-level languages is to port~\SYSTEM's intrinsic functions into the language of choice.

\SYSTEM's verification pipeline proceeds as follows.
First, the user annotates, via functions intrinsic to~\SYSTEM, which variables are universally quantified (e.g., the mechanism inputs and output) and which are probabilistic, along with the originating distribution.
This annotated implementation is then passed to~\SYSTEM alongside the desired level of privacy $\e$.
\SYSTEM then compiles the program to LLVM bitcode and performs probabilistic symbolic execution via~\klee (\Cref{sec:pse}) to collect path conditions and probability expressions for all reachable paths.
These are simplified and transformed into a logical query in SMT-LIB~\citep{smtlib-2.6}, using the Core, FixedSizeBitVectors, FloatingPoint, and ArraysEx theories.
The query is then discharged either by Z3~\citep{demoura2008} or by the exhaustive enumeration algorithm of~\Cref{sec:solver_enum} via~\SYSTEM's SMT-LIB evaluator.
On termination,~\SYSTEM either reports that the implementation satisfies $\e$-differential privacy or returns a concrete counterexample: two adjacent inputs, a mechanism output, and the offending probabilities $\Pr[\M(x) = z]$ and $\Pr[\M(y) = z]$.
\SYSTEM can verify $\e$-differential privacy for all adjacent inputs $(x,y)\in\Delta$, for all adjacent inputs to a fixed concrete input (e.g., $\forall y . (0.5, y) \in \Delta$), or for a single concrete pair (e.g., $(0.0, 1.0)$).

\subsection{Floating-Point Extension to~\klee}
\label{sec:fpklee}

In the process of developing~\SYSTEM, we extended~\klee to support programs that use IEEE~754 binary floating-point formats, operations, and functions.
Our extension builds on the work of~\citet{liew2017}, who explored adding floating-point support to~\klee; we updated their fork to support LLVM 13.0.1, the AArch64 architecture, and the IEEE~754 half-precision (binary16) floating-point type.
We assume that the primitive arithmetic LLVM IR instructions (e.g., \texttt{fadd}) correspond to IEEE~754 operations, whose semantics in turn are used by the SMT-LIB floating-point theory~\citep{brain2015}.
All IEEE~754 rounding modes are supported, as are accurate representations of special values like $\pm0$, $\pm\infty$, and NaN, though we do not distinguish quiet from signaling NaNs.
This limitation arises because the IEEE~754 standard does not precisely specify the binary encoding of either NaN variant, and the SMT-LIB floating-point theory likewise does not define binary encodings for NaN values.
We instead fix a single NaN encoding for each format $\Farb$.
None of the programs studied use NaN payloads, so this does not affect soundness.
As in~\citet{liew2017},~\klee's internal expression language attempts to match Z3's semantics, since Z3 is the solver~\klee uses to check path feasibility.

\subsection{Exhaustive Enumeration Solver}
\label{sec:impl_exhaustive}

We implemented the exhaustive enumeration solver directly in~\SYSTEM by writing an evaluator for the subset of SMT-LIB we use.
Where possible, native primitive types (e.g., \texttt{u64}, \texttt{f64}) are used in place of arbitrary-precision integers and floating-point numbers to improve performance.
Our evaluator's semantics match Z3's and may differ from those of other solvers.
We validate the evaluator by comparing its results against Z3's on equivalent queries.

Recall that for each concrete adjacent input pair $(x,y) \in \Delta$ and each concrete output $z \in \Farb$, the probabilities $\Pr[\M(x, f(\cdot), \e) = z]$ and $\Pr[\M(y, f(\cdot), \e) = z]$ are computed by counting, for each path $i$, the assignments $(v_1, \ldots, v_k) \in \mathcal{D}_i$ to the probabilistic symbolic variables on that path satisfying $\pc_i$.
This computation is embarrassingly parallel: random samples are independent and paths are disjoint, so both the assignments on a single path and the paths within a mechanism can be evaluated concurrently.
We use the \texttt{rayon} data-parallelism framework to schedule the evaluation across all available cores, with a per-thread evaluation context to avoid redundant allocations of intermediate state.
Computed probabilities are memoized and reused.

\subsection{Validation Testing}
\label{sec:validation}

To gain assurance in the correctness of our implementation, we developed a validation procedure that checks whether the exact output distribution~\SYSTEM computes agrees with the actual output distribution of the program.
Recall that~\SYSTEM takes as input a program encoded as LLVM bitcode in which all random samples are modeled by~\klee-intrinsic functions that introduce fresh probabilistic symbolic variables.
This program is not executable: the intrinsic functions only inform~\klee which program variables are symbolic and which distribution each probabilistic symbolic variable is drawn from.
We can, however, compile an executable version by replacing each intrinsic with an equivalent runtime version that uses a pseudo-random number generator to sample concrete values from the same distribution.

Using the executable version and a fixed, concrete input $x$, we estimate the output distribution by running the program $N$ times and counting occurrences of each output.
For each observed output $z$, we compute $\Pr[\M(x, f(\cdot), \e) = z]$ exactly using the technique of~\Cref{sec:meat}.
We then perform Pearson's $\chi^2$ goodness-of-fit test~\citep{pearson1992}, taking the exact probabilities that~\SYSTEM produces as our null hypothesis.
On every program we tested during development, we observed $p$-values extremely close to 1, giving us high confidence in~\SYSTEM's correctness on our case studies.

\section{Evaluation}
\label{sec:eval}

We evaluate our verification technique on two case studies from the literature: the textbook Laplace mechanism~\citep{dwork2014,dwork2006}, which is not $\e$-differentially private over floating-point arithmetic, and the snapping mechanism~\citep{mironov2012}, an alternative designed to resist Mironov's attack (\Cref{sec:mironov}).
We focus on whether~\SYSTEM can (1) reconstruct Mironov's findings by identifying a concrete counterexample to $\e$-differential privacy for the Laplace mechanism, and (2) prove $\e$-differential privacy for an implementation of the snapping mechanism.
\Cref{sec:methodology} describes our experimental setup and case-study implementations;~\Cref{sec:lm} and~\Cref{sec:sm} present the encodings and verification results for the Laplace and snapping mechanisms, respectively; and we close with a discussion of implementation factors that affect~\SYSTEM's performance.

\subsection{Methodology}
\label{sec:methodology}

We implemented both mechanisms as~\CC programs, performing all floating-point arithmetic in accordance with the IEEE~754 standard using native primitive types (e.g., \texttt{float}, \texttt{double}).
Like other symbolic execution methods,~\SYSTEM requires that some parameters be concretized --- particularly those that dictate memory allocation or loop bounds.
For instance,~\klee has limited support for symbolic array lengths and only considers concrete array sizes during path exploration.
Both mechanisms we evaluate use a constant amount of time and space regardless of any supplied parameter, allowing us to fully explore all feasible paths for both implementations.
Such complete exploration is necessary for verifying $\e$-differential privacy: without it,~\SYSTEM can only compute lower bounds on the probabilities, opening the door to false positives.
All experiments were conducted on a machine equipped with a 32-core 2.8\,GHz processor and 128\,GB of memory running Amazon Linux 2023 (kernel \texttt{6.1.168-203.330}) on AArch64.

Both mechanisms are parameterized by the result of a non-private query $f(x)$ where $f: \mathcal{D} \rightarrow \Farb$ and $x \in \mathcal{D}$, and we assume $f$ has sensitivity $\Delta_f = 1$.
Both also require an implementation of the natural logarithm; we write $\LN(\cdot)$ for any such floating-point implementation.

\subsection{Background: Half-Precision Floating-Point Numbers}
\label{sec:half}

We implemented both mechanisms using half-precision floating-point (binary16) arithmetic~\citep{ieee754-2008}.
As the name suggests, half-precision numbers $\half$ offer 11 bits of precision and occupy 16 bits: 5 for the exponent, 10 for the significand, and 1 for the sign.
We use half-precision over single- or double-precision arithmetic to make the verification task tractable: there are far fewer adjacent inputs $(x,y)\in\Delta$ to check and far fewer values in the joint domain of the probabilistic symbolic variables.
Specifically, the standard uniform sampling procedure yields only $2^{11} = 2048$ uniformly-spaced half-precision floating-point numbers in $[0,1)$~\citep{goualard2022}, compared to $2^{24}$ for single precision and $2^{53}$ for double precision.
This shrinks the bitvector terms in~\Cref{eq:bv_dpquery}.
Although binary16 is smaller and less precise, it retains the numeric properties exploited in Mironov's attack (\Cref{sec:mironov}), such as non-uniform spacing and rounding error, and we believe the results in this section transfer to higher-precision formats.

Half-precision arithmetic is natively supported on ARMv8, RISC-V, and some x86 architectures.
Our~\CC implementations use the \texttt{_Float16} type defined by ISO/IEC TS 18661-3:2015~\citep{ISO18661-3:2015}, which implements the IEEE~754-2008 binary16 format~\citep{ieee754-2008}.
Half-precision arithmetic has also seen increasing use in machine-learning workloads~\citep{micikevicius2018,zhao2021,yun2024:arXiv} due to growing GPU support and a favorable size-precision tradeoff, so verification at this width is of independent interest.

\subsection{Laplace Mechanism}
\label{sec:lm}

\begin{algorithm}[t]
  \caption{$\M_{\Lap}$: the half-precision Laplace mechanism with sensitivity 1}
  \label{alg:lm}
  \begin{algorithmic}[1]
    \Function{$\M_{\Lap}$}{$x : \half$}
      \Let{$U$}{$\mathsf{UniformInt}(0, 2^{11}-1) / 2^{11}$}\Comment{\citep{goualard2022}}
      \Let{$r$}{$U \ominus 0.5$}
      \Let{$Y$}{$\mathrm{sign}(r) \otimes \LN(1 \ominus (2 \otimes \abs{r}))$}
      \State \Return{$x \oplus Y$}
    \EndFunction
  \end{algorithmic}
\end{algorithm}

Our first case study is the classic Laplace mechanism (\Cref{sec:background_laplace}); pseudocode is given in~\Cref{alg:lm}.
The mechanism first samples a uniform floating-point number $r \in [-0.5, 0.5)$ by drawing a uniform integer from $[0, 2^{11}-1]$, dividing it by $2^{11}$, and subtracting $0.5$.
It then applies the inverse transform formula $\mathrm{sign}(r)\otimes\ln(1 \ominus (2\otimes\abs{r}))$ before adding the result to the input $x$.
This is~\Cref{eq:lapdist2} instantiated with binary16 floating-point operators.
Assuming sensitivity 1,~\Cref{alg:lm} theoretically satisfies $1$-DP, but we expect our half-precision implementation to be susceptible to the attack of~\citet{mironov2012}.

\begin{table*}[t]
  \setlength{\tabcolsep}{10pt}
  \centering
  \caption{Performance of~\SYSTEM when attempting to verify that $\M_{\Lap}$ satisfies $1$-differential privacy on concrete adjacent inputs $x = +0$ and $y = 1$, across three natural logarithm implementations.
    Counterexamples were produced for every $\LN$ implementation.
    For each $\LN$, we report the number of feasible paths and the total wall-clock time (in seconds) to perform symbolic execution, generate the query, and discharge it via direct SMT discharge (Direct) or exhaustive enumeration (Exhaustive).}
  \label{tab:lm}
  \begin{tabular}{@{}lrrr@{}}
    \toprule
    &&\multicolumn{2}{c}{Timing (sec.)} \\ \cmidrule{3-4}
    \textbf{$\LN$} & \textbf{Paths} & \textbf{Direct} & \textbf{Exhaustive}\\ \midrule
    \approxln & 1 & 6 & 20 \\
    \coremath & 2 & 2563 & 26 \\
    \rlibm & 13 & 2705 & 40 \\
    \bottomrule
  \end{tabular}
\end{table*}

\Cref{tab:lm} reports our experimental results from attempting to verify $1$-differential privacy for the Laplace mechanism.
The adjacent inputs $(x,y) \in \Delta$ were concretized to $x = +0$ and $y = 1$, and a single integer sample was drawn in each variant.
Every experiment found a counterexample consisting of an output value $z \in \{\M_{\Lap}(x') \mid x' \in \half\}$ and two probabilities $\Pr[\M_{\Lap}(0) = z]$ and $\Pr[\M_{\Lap}(1) = z]$ where exactly one of them is zero.

These results corroborate~\citet{mironov2012}: the textbook Laplace mechanism does \textit{not} retain $\e$-differential privacy over floating-point arithmetic.
Moreover,~\SYSTEM produced ``smoking gun'' evidence relatively quickly, in the form of a concrete counterexample equipped with exact probabilities.
\textbf{To the best of our knowledge, this is the first time that the Laplace mechanism implemented in floating-point arithmetic has been formally shown not to be $\e$-differentially private.}

\subsection{Snapping Mechanism}
\label{sec:sm}

\begin{algorithm}[t]
  \caption{$\sm$: the half-precision snapping mechanism with global bound $B$ and privacy parameter $\e$.}
  \label{alg:sm}
  \begin{algorithmic}[1]
    \Function{$\textsc{SampleUlp}$}{}
      \Let{$\mathrm{exponent}$}{$0$}
      \Let{$\mathrm{significand}$}{$0$}
      \While{$\mathrm{exponent} = 0 \wedge \mathrm{significand} = 0$}
        \Let{$\mathrm{exponent}$}{$15 - \textsc{TruncGeom}(0.5, 15)$} \Comment{$\emax = 15$}
        \Let{$\mathrm{significand}$}{$\textsc{UniformInt}(0, 2^{10}-1)$}
      \EndWhile
      \Let{$\mathrm{res}$}{$(\mathrm{exponent} \ShiftLeft 10) \BitOr \mathrm{significand}$}
      \State \Return{$\textsc{InterpretBits}(\mathrm{res})$}
    \EndFunction
    \Function{$\textsc{NextPowerOfTwo}$}{$g: \half$}
      \Let{$\mathrm{significand}$}{$\textsc{ExtractSignificand}(g)$}
      \If{$\mathrm{significand} = 0$}
        \State \Return{$g$}
      \Else
        \Let{$\mathrm{exponent}$}{$\textsc{ExtractExponent}(g) + 1$}
        \State \Return{$\textsc{InterpretBits}(\mathrm{exponent} \ShiftLeft 10)$}
      \EndIf
    \EndFunction
    \Function{$\sm$}{$x : \half$, $B : \half$, $\e : \half$}
      \Let{$U$}{$\textsc{SampleUlp}()$}
      \Let{$\mathrm{sign}$}{$2 \cdot \textsc{UniformInt}(0,1) - 1$} \Comment{$\mathrm{sign} \in \{-1, +1\}$}
      \Let{$\lambda$}{$1.0 \oslash \e$} \Comment{assuming $\Delta_f = 1$}
      \Let{$\widetilde{x}$}{$\textsc{clamp}(x, B) \oplus (\mathrm{sign} \otimes \lambda \otimes \LN(U))$}
      \Let{$\Lambda$}{$\textsc{NextPowerOfTwo}(\lambda)$}
      \Let{$\widehat{x}$}{$\textsc{RoundToMultiple}(\widetilde{x}, \Lambda)$}
      \State \Return{$\textsc{clamp}(\widehat{x}, B)$}
    \EndFunction
  \end{algorithmic}
\end{algorithm}

The snapping mechanism was originally proposed by Mironov~\citep{mironov2012} as an alternative to the Laplace mechanism that retains $\e$-DP under floating-point arithmetic.
As discussed in~\Cref{sec:mironov}, the primary reason the attack against the Laplace mechanism succeeds is that sampling from the Laplace distribution yields a very porous distribution that misses many representable values.
These distributions are not translation-invariant, so after shifting the sample by $f(x)$ and $f(y)$ for adjacent inputs $(x,y) \in \Delta$, the two distributions diverge significantly.
This phenomenon persists regardless of the precision of the underlying floating-point format or the numerical accuracy of the output: due to the non-uniformity of sampling from $(0,1)$ and to rounding error, some outputs can only be produced from $x$ and others only from $y$, enabling an adversary to distinguish the input by observing the output.
The snapping mechanism avoids this by making the set of possible output values independent of the input and uniformly spacing them by approximately $\Delta_f / \e$.

Pseudocode for our implementation $\sm$ is given in~\Cref{alg:sm}.
It is heavily influenced by two publicly available double-precision implementations of the snapping mechanism~\citep{covington2019,diffprivlib}.
The mechanism takes three half-precision inputs: the result of some non-private query $x$, a symmetric global bound on the outputs $B$, and the desired privacy parameter $\e$.
All outputs are restricted to $[-B, B]$.
Following~\citet{mironov2012}, the first step is to draw from the uniform distribution over $\half \cap (0,1)$ where each half-precision floating-point number is output with probability proportional to its unit-in-the-last-place (ULP).
As hinted at by Mironov~\citep{mironov2012}, we implement this by independently sampling an exponent from a truncated\footnote{We define the probability mass function of $\textsc{TruncGeom}(p, n)$ as $\Pr[X = k] =
  \begin{cases}
    (\frac{1}{2})^{k+1}, & \text{if } 0 \leq k < n, \\
    (\frac{1}{2})^{k}, & \text{if } k = n.
  \end{cases}
  $} geometric distribution $\textsc{TruncGeom}(\frac{1}{2}, 15)$ and a significand by uniformly drawing a bitstring from $\{0,1\}^{10}$.
A half-precision number is then constructed by shifting the exponent into place and OR-ing it with the significand; $\textsc{InterpretBits}(\cdot)$ reinterprets the bits of the resulting machine integer as a half-precision floating-point number.
A careful reader may notice that $\textsc{SampleUlp}$ contains an infinite trace in which 0 is repeatedly drawn for both the exponent and significand.
We handle this edge case in~\SYSTEM by evenly distributing the probability mass of $+0.0 \in \half$ across all other possible outputs, allowing us to explore all feasible paths.
The sign bit is sampled uniformly from $\{-1, +1\}$.

Once the samples are drawn, the snapping mechanism computes
\[
  \widetilde{x} = \textsc{clamp}(x, B) \oplus (\mathrm{sign} \otimes \lambda \otimes \LN(U)),
\]
where $\textsc{clamp}(x, B)$ returns $B$ if $x > B$, $-B$ if $x < -B$, and $x$ otherwise.
Mironov requires $\LN$ to be correctly rounded~\citep{mironov2012}.
Next, $\textsc{NextPowerOfTwo}(\cdot)$ computes $\Lambda$, the smallest (possibly negative) power of 2 such that $\lambda \leq \Lambda$, avoiding additional floating-point error.
$\textsc{RoundToMultiple}(\widetilde{x}, \Lambda)$ then rounds $\widetilde{x}$ to the closest multiple $\widehat{x} \in \half$ of $\Lambda$, with ties resolved toward $+\infty$, by directly manipulating the binary representation of $\widetilde{x}$ to avoid rounding artifacts.
The mechanism concludes by returning $\widehat{x}$ clamped to $[-B, B]$.

According to~\citet[Theorem 1]{mironov2012}, the double-precision variant of the snapping mechanism satisfies $(\e + 2^{-49}B\e)$-differential privacy when $1/\e < B < 2^{46}/\e$.
We adapt Mironov's analysis to half-precision by following his proof, substituting the binary16 machine epsilon $\eta = 2^{-10}$ wherever the original analysis uses the double-precision machine epsilon of $2^{-53}$.
\begin{theorem}
  \label{thm:sm}
  The half-precision variant of the snapping mechanism $\sm$, as defined in~\Cref{alg:sm} and assuming sensitivity $\Delta_f = 1$, satisfies $(\e + 2^{-6}B\e)$-differential privacy when $\e < B < 2^4\e$.
\end{theorem}

\begin{table*}[t]
  \setlength{\tabcolsep}{4pt}
  \centering
  \caption{Performance of~\SYSTEM when attempting to verify that $\sm$ satisfies $\widehat{\e}$-differential privacy on concrete adjacent inputs $x = +0$ and $y = 1$, across three settings of the privacy parameters and three natural logarithm implementations.
    $\widehat{\e}$-differential privacy was verified in every experiment.
    Each row reports, for a given natural logarithm implementation $\LN$ and privacy parameters $\e$ and $B$, the privacy level $\widehat{\e}$ guaranteed by~\Cref{thm:sm}, the number of feasible paths, the approximate spacing $\lambda$ between adjacent elements of the output domain, the cardinality of the output domain (\#~Outputs), and the total wall-clock time (in seconds) to perform symbolic execution, generate the query, and discharge it via direct SMT discharge (Direct) or exhaustive enumeration (Exhaustive).}
  \label{tab:sm}
  \begin{tabular}{@{}lrrrrrrrr@{}}
    \toprule
    &&&&&&&\multicolumn{2}{c}{Timing (sec.)} \\ \cmidrule{8-9}
    \textbf{$\LN$} & $\e$ & $B$ & $\widehat{\e}$ & \textbf{Paths} & $\lambda$ & \textbf{\# Outputs} & \textbf{Direct} & \textbf{Exhaustive}\\ \midrule
    \approxln & 1.0 & 10.0 & 1.06 & 13 & 1.0 & 22 & 264 & 49 \\
    \coremath & 1.0 & 10.0 & 1.06 & 15 & 1.0 & 22 & 160 & 81 \\
    \rlibm & 1.0 & 10.0 & 1.06 & 15 & 1.0 & 22 & 1655 & 455 \\\midrule
    \approxln & 2.0 & 1.75 & 2.021 & 13 & 0.5 & 10 & 196 & 33 \\
    \coremath & 2.0 & 1.75 & 2.021 & 15 & 0.5 & 10 & 142 & 55 \\
    \rlibm & 2.0 & 1.75 & 2.021 & 15 & 0.5 & 10 & 1928 & 294 \\\midrule
    \approxln & 4.0 & 1.125 & 4.027 & 13 & 0.25 & 12 & 264 & 36 \\
    \coremath & 4.0 & 1.125 & 4.027 & 15 & 0.25 & 12 & 133 & 54 \\
    \rlibm & 4.0 & 1.125 & 4.027 & 15 & 0.25 & 12 & 1620 & 305 \\
    \bottomrule
  \end{tabular}
\end{table*}

\Cref{tab:sm} summarizes our experimental results from attempting to verify $\e$-differential privacy for the snapping mechanism across three settings of the privacy parameters $\e$ and $B$.
As with the Laplace mechanism, we vary both the natural logarithm implementation and the solver strategy.
The adjacent inputs $(x,y) \in \Delta$ were concretized to $x = +0$ and $y = 1$, and each program sampled three random values.

\textbf{\SYSTEM verified differential privacy in every experiment above. To the best of our knowledge, this is the first time the snapping mechanism has been verified to satisfy differential privacy over floating-point arithmetic.}
We were also able to prove $\widehat{\e}$-DP across several settings of $\e$ and $B$, each inducing a different (approximate) spacing between elements of the output domain.
Notably,~\SYSTEM verified $\widehat{\e}$-DP even when using the~\approxln implementation of $\LN$, which is not known to be correctly rounded for half-precision; we hypothesize this is because~\approxln computes the natural logarithm in single-precision rather than half-precision arithmetic.

\textbf{Overall, the exhaustive enumeration strategy was substantially faster than direct discharge, for two reasons.}
First, exhaustive enumeration is embarrassingly parallel and takes full advantage of the threads on our evaluation machine, whereas Z3 runs as a single-threaded process.
Second, the queries provided to Z3 are semantically complex.
The floating-point and bitvector arithmetic embedded in the path conditions is highly non-linear, partly because we sample uniform floating-point numbers by dividing two integers.
Worse, much of~\Cref{alg:sm} and all three $\LN$ implementations consist of bitwise operations on the underlying bitvector encoding a floating-point number (its binary interchange format).
Converting floating-point numbers into their binary interchange format is not officially supported by SMT-LIB's floating-point theory~\citep{brain2015} as it is not a total function: SMT-LIB has a single NaN per floating-point format, while many bitvectors encode NaN under IEEE~754.
Z3's \texttt{fp.to_ieee_bv} performs this conversion but leaves the NaN payload unconstrained, exploding the search space.
Since we do not care about NaN payloads, we fix the binary interchange representation of NaN in each format $\Farb$, yielding a total function between $\half$ and 16-bit bitvectors.
\textbf{Even so, these results suggest that SMT solving over the floating-point and bitvector theories is the primary performance bottleneck.}

\subsection{Correctly-Rounded Natural Logarithm}
\label{sec:log}

Both mechanisms in our evaluation reduce to a single floating-point evaluation of the natural logarithm followed by a small amount of additional arithmetic, so the choice of $\LN$ implementation directly determines the support of the output distribution.
Mironov's snapping mechanism requires $\LN$ to be correctly rounded~\citep{mironov2012}: a floating-point function $\widehat{f}: \Farb \to \Farb$ is \emph{correctly rounded} with respect to a real-valued function $f: \mathbb{R} \to \mathbb{R}$ if, for every $x \in \Farb$, $\widehat{f}(x)$ equals the floating-point number obtained by rounding $f(x)$ under the active rounding mode~\citep{muller2018}.
Correctly-rounded transcendental functions are notoriously difficult to construct due to the table maker's dilemma, and standard math library implementations (including \textsc{glibc}'s) typically lack any rounding guarantees.

We evaluate~\SYSTEM against three different $\LN$ implementations, two of which are known to be correctly rounded.
The~\rlibm research project~\citep{lim2022} provides correctly rounded elementary functions for multiple representations (e.g., binary16, binary32, bfloat16, tensorfloat32) and multiple rounding modes.
The~\coremath project~\citep{sibidanov2022} is a more recent successor to the now-abandoned CRlibm~\citep{daramy-loirat2006} project, supplying open-source correctly rounded elementary functions for the binary16, bfloat16, binary32, and binary64 formats across all rounding modes.
Both implementations store polynomial coefficients in lookup tables and therefore require SMT-LIB's \texttt{ArraysEx} theory to model.

Our third implementation,~\approxln, uses the classical range-reduction method of~\citet{cody1980,defour2001} and a minimax low-degree polynomial approximation of the natural logarithm.
It is \textit{not} correctly rounded, is implemented for single-precision, uses no lookup tables, and consists mostly of straight-line code.
We use it on binary16 inputs by widening to binary32 before the call and narrowing the result back to binary16 afterward.

The timings in~\Cref{tab:lm} and~\Cref{tab:sm} strongly suggest that the choice of $\LN$ has serious implications for~\SYSTEM's performance.
For both mechanisms,~\SYSTEM performed significantly better on~\approxln than on~\rlibm or~\coremath.
We attribute this to the large lookup tables in~\rlibm and~\coremath, which make path conditions more complex to reason over, require SMT-LIB's \texttt{ArraysEx} theory, and increase the query size.
Between~\coremath and~\rlibm,~\coremath performed better, likely because it has fewer branches and is implemented in half-precision arithmetic, whereas~\rlibm uses a mix of single- and double-precision arithmetic.
The choice of $\LN$ implementation did not affect the verification result in any of our experiments.

\section{Related Work}
\label{sec:rw}

\paragraph*{Verification and testing for differential privacy.}

Differential privacy is a practically important and challenging example of a
probabilistic property, and researchers have developed numerous methods to
ensure that programs satisfy differential privacy. There are now verification
approaches based on domain-specific languages, type systems, and deductive
verification, which are too numerous to survey here (e.g.,
\citep{mcsherry2009,reed2010,barthe2013,barthe2015,DBLP:journals/pacmpl/MedeirosNLKRZJTAT25});
the interested reader can consult a recent survey~\citep{Murawski:2016:2893582}.
While these methods provide formal guarantees, they all tackle simplified core
languages with idealized behavior, such as continuous random sampling, or random
sampling over the (unbounded) integers. Existing verification works do not
consider programs with floating-point arithmetic. In contrast, our method
tackles real implementations in existing programming languages.

A more recent line of work finds counterexamples to differential
privacy~\citep{bichsel2018,bichsel2021}. Notably, the method by
\citet{bichsel2021} can be applied given only black-box access to the program,
and can be used to detect violations of differential privacy in real
implementations. These methods rely on executing the program and performing
random sampling, and provide a statistical guarantee. In contrast, our method
computes probabilities exactly, and provides logical guarantees.

\paragraph*{Differential privacy and floating-point effects.}

By exhibiting attacks on differential privacy, Mironov~\citep{mironov2012}
showed that there is a significant gap between the differential privacy
guarantees of idealized private algorithms using continuous sampling, and
concrete implementations using floating-point sampling. Mironov proposed the
snapping mechanism as a differentially private algorithm using floating-point
sampling. Our method is the first to formally verify any kind of privacy
guarantee for this intricate algorithm. \citet{ilvento2020} has proposed
slightly modifying the definition of differential privacy itself in order to
ease practical implementation. Finally, floating-point arithmetic can compromise
differential privacy through timing side channels~\citep{DBLP:conf/sp/AndryscoKMJLS15},
which are not considered by our method.

\paragraph*{Symbolic execution methods.}

Our method relies on symbolic execution, a verification technique with a long
history~\citep{king1976}. We build on work by~\citet{susag2022} that develops a
symbolic execution method for probabilistic programs with discrete sampling;
more recent work considers symbolic execution for programs with continuous
sampling~\citep{voogd2023}. We also build on~\klee~\citep{cadar2008}, a mature
and industrial-strength symbolic execution engine, along with recent extensions
to handle programs with floating-point arithmetic~\citep{liew2017}.

\section{Conclusion and Future Directions}
\label{sec:conclusion}

We presented a symbolic-execution-based automatic verifier of $\e$-differential privacy for randomized programs implemented in IEEE~754 floating-point arithmetic.
We implemented our technique in a tool called~\SYSTEM and evaluated it on two case studies: the Laplace mechanism, for which it reconstructs Mironov's attack by finding a concrete counterexample, and the snapping mechanism, for which it provides what is, to our knowledge, the first automated proof of $\e$-differential privacy for any concrete implementation.

One natural direction for future work is improving verification performance.
Our exhaustive enumeration strategy already substantially outperforms direct SMT discharge, but the per-path probability formulas remain large and SMT solving over the floating-point and bitvector theories is a clear bottleneck.
Knowledge-compilation techniques such as binary decision diagrams could compactly represent the per-path probability expressions and amortize work across paths.
Another possibility is to develop better methods for simplifying path probability expressions, potentially using equality saturation.
Tangentially, improving SMT solver support for the floating-point theory --- or developing decision procedures targeted at the fragments that arise from probabilistic symbolic execution --- would benefit both strategies.

Another direction is extending~\SYSTEM to support other variants of differential privacy and mechanisms.
For example,~\citet{ilvento2020} proposed a base-2 reformulation of the exponential mechanism, designed to sidestep transcendental rounding.

%
%
%
%

\bibliography{bibliography}
\end{document}